\documentstyle[pra,aps,twocolumn]{revtex}
\begin{document}
\draft

\twocolumn[\hsize\textwidth\columnwidth\hsize\csname @twocolumnfalse\endcsname

\title{An SO(5) model of p-wave superconductivity and ferromagnetism}
\author{Shuichi Murakami and Naoto Nagaosa}
\address{Department of Applied Physics, University of Tokyo,
Bunkyo-ku, Tokyo 113-8656, Japan}
\author{Manfred Sigrist}
\address{Yukawa Institute for Theoretical Physics, Kyoto University,
Kyoto 606-8502, Japan}
\date{\today}
\maketitle
\begin{abstract}
We study an SO(5) model unifying p-wave superconductivity and ferromagnetism. 
If only a single p-wave pairing wavefunction 
is involved, the p-wave superconducting
(pSC) and ferromagnetic (F) states can be unified to form 10-dimensional 
multiplet of SO(5). 
The collective modes in pSC and F states are studied from this viewpoint 
in terms of a non-linear sigma model.
\end{abstract}

\vskip2pc]

\narrowtext
Proximity of magnetism and superconductivity is one of the most 
important issues in the physics of strongly correlated electronic systems.
In heavy fermion systems \cite{hv1,hv2}, 
organic conductors \cite{org1},  and 
high-$T_{c}$ superconductors(HTSC) \cite{htsc}, 
the interplay between antiferromagnetism (AF) and anisotropic 
superconductivities (SC) has been 
studied. Recently, a theory has been proposed to unify AF and 
d-wave SC (dSC) in terms of the SO(5) symmetry~\cite{zhang}. 
In this theoretical 
framework, the order parameters of AF and dSC form a five dimensional 
supervector $\vec n$. 
The formation of the pseudo-gap  is identified with the 
onset of the amplitude of this supervector $| \vec n|$.
The SO(5) rotation in this five dimensional space turns AF 
into dSC and vice versa. 
A prediction of this theory is a specific collective mode in the
superconducting state, so-called 
$\pi$-mode~\cite{demler}, which corresponds to the fluctuation towards 
the AF. The corresponding $\pi$-operator becomes canonically
conjugate to the AF spin fluctuation 
below the superconducting transition and, hence, can be detected by 
neutron 
scattering experiments. This gives a possible scenario for the 
41$meV$-peak observed in YBa$_2$CuO$_{6.9}$~\cite{exp1,exp2,exp3,exp4,exp5}.
Furthermore, this unification is supported by some of the recent 
numerical studies~\cite{mhdz,ehz,headmz}. 

Let us now turn to p-wave superconductors.
Ferromagnetic spin fluctuations have long been known to mediate the 
p-wave pairing in $^3$He~\cite{leggett}. 
A more recent example is Sr$_2$RuO$_4$~\cite{maeno}.
This compound shows a superconducting state below $T_{c}\sim 1.5{\rm K}$.
Unconventional pairing is clearly indicated by the 
high sensitivity of $T_{c}$ to nonmagnetic 
impurities~\cite{mackenzie}, and the absence 
of Hebel-Slichter peak in NQR spectrum~\cite{ishida1}.
Theoretically it was proposed to be a two-dimensional p-wave 
superconductor~\cite{rs}, and this was confirmed by ${}^{17}$O 
NMR Knight shift measurement~\cite{ishida2}
showing the temperature-independent 
behavior of in-plane susceptibility.
Among the five p-wave pairing states suggested from group theoretical
arguments~\cite{rs},   
$\vec{d}(\vec{k})=\hat{z}(k_{x}\pm{\rm i}k_{y})$ is the 
only order parameter~\cite{xiy1}
compatible with broken time-reversal symmetry 
observed by $\mu$SR experiments~\cite{xiy2}.
Interestingly, this compound is close to ferromagnetic order in the 
following aspects. First, NMR data indicate the enhancement of
ferromagnetic spin fluctuations~\cite{ferro}. Second, high pressure
destroys superconductivity and drives the material towards
ferromagnetism~\cite{yoshida}. Above 3 GPa the superconducting state
has disappeared and around 8 GPa  
the in-plane resistivity is proportional to $T^{4/3}$ consistent with 
the behavior dominated by two-dimensional ferromagnetic
spin fluctuations.
Lastly, among related materials Sr$_{n+1}$Ru$_{n}$O$_{3n+1}$, 
ferromagnetic order is observed for
$n=(2),3,\infty$~\cite{gibb,neumeier,cao}, 
and with increasing $n$ the ferromagnetic transition temperature 
rises.

Motivated by this, we study in this paper an SO(5) model 
describing the interplay between ferromagnetism (F) and p-wave
superconductivity (pSC). The charge $Q$, the order parameter of pSC,
and the total spin $\vec S$ constitute a ten-dimensional superspin
obeying the closed SO(5) algebra. Note that this algebra was
found in the context of $^3$He superfluidity~\cite{leggett1,ha1,ha2}.
We consider the case where the 
amplitude of this superspin is developed and the 
rotational fluctuations between F and pSC states constitute low-lying
excitations.

In the weak-coupling BCS theory one introduces a gap function, 
$
\Delta_{\alpha\beta}({\bf k})=-\sum_{{\bf k}',\gamma,\delta}
V_{\alpha\beta\gamma\delta}
({\bf k},{\bf k}')\langle c_{-{\bf k}'\gamma} c_{{\bf k}'\delta}\rangle,
$
where $V({\bf k},{\bf k}')$ represents the electron-electron 
interaction.
In this paper we focus on odd-parity pairing, where
the gap function can be expressed as 
$
\hat{\Delta}({\bf k})=
\sum_{\mu}d_{\mu}({\bf k}){\rm i}{\sigma}^{\mu}
{\sigma}^{y}.
$
If we assume that $V({\bf k},{\bf k}')$ is nonzero only 
when ${\bf k}$ and ${\bf k}'$ are close to the Fermi surface,
$V({\bf k},{\bf k}')$ 
depends only on the directions $\hat{{\bf k}}$ and $\hat{{\bf k}}'$.
Hence the order parameter of spin-triplet SC state is 
\begin{equation}
d_{\mu}(\hat{{\bf k}})=
{\rm i}
\sum_{{\bf p},\alpha,\beta}
V^{(t)}(\hat{{\bf k}},\hat{{\bf p}})
\langle c_{-{\bf p}\alpha} c_{{\bf p}\beta}\rangle
({\sigma}^{y}
{\sigma}^{\mu})_{\alpha\beta}
\end{equation}
where $V^{(t)}$ is the odd-parity part of $V({\bf k},{\bf k}')$, i.e.
$V^{(t)}(\hat{{\bf k}},\hat{{\bf p}})$ is an odd function in both 
$\hat{{\bf k}}$ and $\hat{{\bf p}}$. 
The summation over ${\bf p}$ is restricted to the region close to the 
Fermi surface. 
The lowest order term in the expansion of 
$V^{(t)}(\hat{{\bf k}},\hat{{\bf p}})$ in terms of spherical harmonics  yields
$V^{(t)}(\hat{{\bf k}},\hat{{\bf p}})
\propto\hat{{\bf k}}\cdot\hat{{\bf p}}$ ,
which corresponds to the pSC order parameter ($l=1$).

We now introduce a new operator
\begin{equation}
\pi_{\mu}(\hat{{\bf k}})={1 \over 4}\sum_{{\bf p},\alpha,\beta}
g(\hat{{\bf k}},\hat{{\bf p}})c_{-{\bf p},\alpha}
({\sigma}^{y}{\sigma}^{\mu})_{\alpha\beta}c_{{\bf p}\beta},
\end{equation}
with
$g(\hat{{\bf k}},\hat{{\bf p}})\propto V^{(t)}(\hat{{\bf k}},
\hat{{\bf p}})$.
It obviously satisfies
$\langle\pi_{\mu}(\hat{{\bf k}})\rangle\propto d_{\mu}(\hat{{\bf k}})$.
The $\hat{{\bf k}}$-dependence indicates angular dependence of 
the pair wavefunction in 
the pSC phase. At this point we may recall that 
in SO(5) theory 
of the HTSC, the pair wavefunction is restricted to 
$d_{x^2-y^2}$ \cite{zhang}.
A similar simplification is expected, if the rotational symmetry in 
the orbital space is reduced to the discrete point group (tetragonal
for Sr$_2$RuO$_4$).
Therefore, we consider from now on a definite form of the pair 
wavefunction~\cite{multi},
which can be implemented by fixing the $\hat{{\bf k}}$-dependence of $ 
V$, which allows us to suppress the $\hat{{\bf k}}
$-dependence as $g(\hat{{\bf p}})=
g(\hat{{\bf k}},\hat{{\bf p}})$ unless necessary.
Therefore, superfluid ${}^{3}$He is outside the scope of this SO(5) theory,
since the spatial part of the pair wavefunction is not fixed.

Let us define the SO(5) generators $L_{ab}$ of rotations in the 
F-pSC space as elements of an
antisymmetric $ 5 \times 5 $-matrix
\begin{equation}
\left(
  \begin{array}{ccccc}
0 &&&&\\
\pi_{x}^{\dagger}+\pi_{x} & 0 &&& \\
\pi_{y}^{\dagger}+\pi_{y} & -S_{z} &0&& \\
\pi_{z}^{\dagger}+\pi_{z} & S_{y} &-S_{x}&0 & \\
Q & 
{\rm i}(\pi_{x}-\pi_{x}^{\dagger}) &
{\rm i}(\pi_{y}-\pi_{y}^{\dagger}) &
{\rm i}(\pi_{z}-\pi_{z}^{\dagger}) &0
  \end{array}
\right)
\label{eqn:L}
\end{equation}
where 
$Q={1\over 2}\sum_{{\bf p},\alpha} \left( c_{{\bf p}\alpha}^{\dagger}
c_{{\bf p}\alpha}-{1\over 2}\right)$, and $\ S_{\mu}=
{1\over 2}\sum_{{\bf p},\alpha,\beta} c_{{\bf p}\alpha}^{\dagger}
{\sigma}^{\mu}_{\alpha\beta}c_{{\bf p}\beta}.$
Provided
$|g(\hat{{\bf p}})|^{2}=1$,
the generators $L_{ab}$ satisfy SO(5) commutation 
rules~\cite{leggett1,ha1,ha2}: 
\begin{equation}
[L_{ab},L_{cd}]
={\rm i}(\delta_{ac}L_{bd}-\delta_{ad}L_{bc}
-\delta_{bc}L_{ad}+\delta_{bd}L_{ac}).
\label{eqn:LL}
\end{equation}
Considering that $g(\hat{{\bf k}}, \hat{{\bf p}})$ 
should be an odd function of 
$\hat{{\bf k}}$ and $\hat{{\bf p}}$, we see that 
$|g|^{2}=1$ exactly 
holds with the choice $
g(\hat{{\bf k}},\hat{{\bf p}})=\mbox{sgn}(
\hat{{\bf k}}\cdot\hat{{\bf p}}) $,
with fixed value of $\hat{{\bf k}}$.
There are other choices, e.g. $ g(\hat{{\bf p}})=
(\hat{p}_{x}
\pm {\rm i}\hat{p}_{y})/\sqrt{2} $, 
which is possibly realized in Sr$_{2}$RuO$_{4}$~\cite{xiy1,xiy2}.

We now define a superspin $\vec{n}$ of SO(5), which 
can be conveniently labeled by two indices as $n_{ab}$.
There is some freedom to define $n_{ab}$, and the simplest choice is 
$n_{ab}=L_{ab}$. Instead, we can also define $n_{ab}$ similarly to 
$L_{ab}$, but with an extra factor $w_{{\bf p}}$~\cite{rkdz} 
inside the summation over ${\bf p}$ in each operator, satisfying 
$w_{{\bf p}}=w_{-{\bf p}}$.
A transformation rule of the superspin $n_{ab}$ under SO(5) is 
determined by the commutation relation
\begin{equation}
[L_{ab},n_{cd}]
={\rm i}(\delta_{ac}n_{bd}-\delta_{ad}n_{bc}
-\delta_{bc}n_{ad}+\delta_{bd}n_{ac}).
\label{eqn:Ln}
\end{equation}
There is an obvious similarity between (\ref{eqn:LL}) 
and (\ref{eqn:Ln}) indicating 
that the commutation relation (\ref{eqn:Ln}) corresponds to the 
ten-dimensional adjoint representation.
In contrast, in the SO(5) theory of HTSC~\cite{zhang},
the superspin is five-dimensional, constituting the vector representation.
Here we shall set $w_{\bf p}=1$, i.e. 
$n_{ab}=L_{ab}$ to assure that $L_{34},L_{42},L_{23}$ are the spin 
operators $\vec{S}$.
Thus, $L_{ab}$ plays both the roles of 
SO(5) generator and superspin.
The physical interpretation of each element of $L_{ab}$ is the 
following: $S_{\mu}$ and $\pi_{\mu}$ are the order parameters of the F 
and pSC phase, respectively, and $Q$ is the electron number.

Now we construct a non-linear sigma model for the superspin $\vec L$
assuming that the length $|\vec L|$ has already been developed
and only the  direction represents low-lying degrees of
freedom.  This appearance of $|\vec{L}|$  occurs, in general, at a temperature higher 
than $T_{c}$. At the present stage it is not possible to identify
clearly the onset of $|\vec{L}|$ in Sr$_2$RuO$_4$, although certain 
crossover behaviors have been seen in various
quantities, e.g. resistance \cite{maeno}. Note, however, that amplitude fluctuations of the order
parameter are irrelevant in the sense of renormalization group in two dimensions.
The detailed analysis of the onset of $|\vec{L}|$ is left for 
a future problem.

The Hamiltonian consists of the anisotropy energy in the superspin space 
and the elastic energy,
i.e.\ spatial rigidity, as
\begin{equation}
H=
\sum_{a<b}\frac{1}{2\chi_{ab}}L_{ab}^{2}+
\sum_{a<b}\frac{\rho_{ab}}{2}(v_{ab})^{2},
\label{eqn:Hamiltonian}
\end{equation}
where  the generalized 
velocity $v_{ab}=\sum_{c}(L_{ac}\nabla L_{bc}-L_{bc}\nabla L_{ac})$.
For simplicity, we set $\rho_{ab}=\rho_{ba}$,  
$\chi_{ab}=\chi_{ba}$.
Gauge invariance and spin rotational invariance fix the form of the first
term in (\ref{eqn:Hamiltonian}), because
neither linear terms nor cross terms in $L_{ab}$ are allowed. 
Since $\vec{L}$ is ten-dimensional, 
it might be possible to include all the rotations in the
ten-dimensional space into the first term, i.e.\ SO(10) generators.
Nevertheless, we consider here only SO(5) generators, because 
it is the minimal symmetry containing both SO(3) spin-rotational symmetry 
of F phase and U(1)-gauge symmetry of pSC.
We do not know so far whether a construction of SO(10) generators in 
a second-quantized form is possible.
It is natural to assume that $\chi_{ab}$ ($\rho_{ab}$) in 
(\ref{eqn:Hamiltonian}) are identical with each other within the 
same phase. Thus, we set 
$\chi_{1a}=\chi_{a5}=\chi_{\pi} \ (a=2,3,4)$, $\chi_{15}=\chi_{Q}$,
$\chi_{23}=\chi_{24}=\chi_{34}=\chi_{S}$ and similarly for $\rho$.
We shall also add by hand one more symmetry-breaking term
\begin{equation}
H_{\pi}=g(\vec{\pi}\times \vec{\pi}^{*}-
\vec{\pi}^{*}\times \vec{\pi})^{2} \ \ \ \ (g<0),
\label{eqn:Hpi}
\end{equation}
which is proportional to $(\vec{d}\times\vec{d}^{*})^{2}$. 
Since the $\vec{d}$-vector $\vec{d}(\vec{k})=\hat{z}(k_{x}\pm{\rm i}k_{y})$ 
realized in Sr$_{2}$RuO$_{4}$ corresponds to a unitary state,
we assume $g<0$ to assign lower energy to unitary states 
($\vec{\pi}\parallel\vec{\pi}^{*}$) in 
pSC than nonunitary states.
This kind of term has appeared previously in Ref.\cite{ab}
in connection with the discussion of the stability of the ABM state in 
a certain region of the  phase diagram of ${}^{3}$He.
In contrast to ${}^{3}$He, where the continuous rotational
symmetry of pair wavefunctions admits five terms of this kind,
here the quenched orbital part allows for
only two terms: $H_{\pi}$ and
$H' = (|\mbox{Re}\vec{d}|^{2}+|\mbox{Im}\vec{d}|^{2})^{2}$.
The latter term contributes to the anisotropy between   
F and pSC states. However this anisotropy has been already taken into account 
in anisotropies of $\chi$'s in the first term of eq.(12)
since we are considering the case 
with fixed $|\vec{L}|$. Therefore, $H'$ does not give qualitatively
new effects.

Let us discuss the parameters in (\ref{eqn:Hamiltonian}). 
An excitation associated with  the term $Q^{2}/(2\chi_{Q})$ corresponds 
to a plasmon mode with high energy. 
Therefore, we set $\chi_{Q}\ll \chi_{S},\chi_{\pi}$.
The ground state is either pSC or F, 
depending on the values of $\chi_{S}$ and $\chi_{\pi}$. 
If $\chi_{\pi}<\chi_{S}$
($\chi_{\pi}>\chi_{S}$)
, the system 
is in the F (pSC) phase.

As in \cite{zhang}, 
one can use Hamiltonian (\ref{eqn:Hamiltonian}) (\ref{eqn:Hpi})
to study collective modes.
The equation of motion is given by 
\begin{eqnarray}
&&\dot{L}_{ab}=
{1\over 2}\sum_{d}
(\chi_{ad}^{-1}-\chi_{bd}^{-1})\{ L_{ad}, L_{bd}\}
+{1 \over 2}\sum_{d}(\rho_{ad}-\rho_{bd})\{ v_{ad},v_{bd}\} \nonumber \\
&&+{1 \over 2}\sum_{c,d}\rho_{ac}\nabla\{
v_{ac}, L_{cd}L_{bd}\}
-{1 \over 2}\sum_{c,d}\rho_{bc}\nabla\{
v_{bc}, L_{cd}L_{ad}\}\nonumber \\
&&+{1 \over 2}\sum_{c,d}\rho_{cd}\nabla\{
v_{cd}, L_{ca}L_{bd}-L_{cb}L_{ad}\}+{\rm i}{[}H_{\pi},L_{ab}{]},
\label{eqn:motion}
\end{eqnarray}
with $\{ A, B \} = AB + BA$. 
Using (\ref{eqn:motion}), we can discuss collective modes 
in each phase: pSC and F.

When $\chi_{\pi}>\chi_{S}$ and $g<0$, all the unitary states 
form a set of degenerate ground states. 
Without loss of generality 
we consider the state with 
$\langle L_{41}\rangle =-\langle 
L_{14}\rangle
=A_{\pi}$ and other $L_{ab}$'s are vanishing. 
When we linearize the equation of motion (\ref{eqn:motion}) around this
ground state,
we get
\begin{eqnarray}
\dot{L_{c1}}&=&(\chi_{\pi}^{-1}-\chi_{S}^{-1})A_{\pi}
L_{c4}+\rho_{\pi}A_{\pi}^{3} \nabla^{2}
L_{c4}  \ \ (c=2,3)
\label{eqn:c1}
\\
\dot{L_{c4}}&=&-\rho_{S}A_{\pi}^{3}\nabla^{2}
L_{c1}  \ \ (c=2,3)
\label{eqn:c2}
\\
\dot{L_{51}}&=&\rho_{Q}A_{\pi}^{3}\nabla^{2}
L_{54}
\label{eqn:51}
\\
\dot{L_{54}}&=&(\chi_{Q}^{-1}-\chi_{\pi}^{-1})L_{51}
-\rho_{\pi}A_{\pi}^{3}\nabla^{2}L_{51} 
\label{eqn:52} \\
\dot{L_{cd}}&=&0 \ \  \ (c,d\in \{2,3,5\})
\label{eqn:cd} 
\end{eqnarray}
At a first glance (\ref{eqn:cd}) seems to represent Goldstone modes, but
it is not the case, because $L_{cd}$ ($c,d\in \{2,3,5\}$)
is not coupled to $L_{41}$. 
On the other hand, the rotations in spin space give rise to Goldstone 
modes, which are described by the 
coupled equations (\ref{eqn:c1}) and (\ref{eqn:c2}).
The commutation relation
$[\pi_{\lambda}, \ {S}_{\mu}]={\rm i}\varepsilon_{\lambda\mu\nu}
\pi_{\nu}$
gives canonical conjugate relation between 
spin and $\pi$'s, when pSC order parameter in the r.h.s.
is replaced by a nonzero expectation value, and
$L_{24} = - S_y$ and $L_{34}= S_x$
correspond to rotations of the $\vec{d}$-vector in pSC phase, 
while $L_{21}= \pi_x+ \pi_x^\dagger$ and 
$L_{31}= \pi_y + \pi_y^\dagger$ to rotations toward the F phase~\cite{ab}.
They yield two collective modes with the frequency
\begin{equation}
\omega(k)=A_{\pi}^{2}k\sqrt{
\rho_{S}
(\chi^{-1}_{S}-\chi_{\pi}^{-1}+\rho_{\pi}A_{\pi}^{2}k^{2})}.
\label{eqn:disp}
\end{equation}
One new aspect here compared with previous discussion \cite{ab}
is that the ``momenta'' $\pi$ correspond to the fluctuation 
toward F state and the frequency is determined by 
the anisotropy energy in superspin space, 
$\chi^{-1}_S-\chi^{-1}_\pi$.  
Lastly, the mode in (\ref{eqn:51}) and (\ref{eqn:52}) can be neglected,
because they move up to 
high energies when long range 
Coulomb interaction is considered.

When the system is in the F phase ($\chi_{\pi}<\chi_{S}$), 
we linearize (\ref{eqn:motion}) around the ground state corresponding to 
$ \langle L_{23} \rangle = \langle S_{z} \rangle = A_S$ 
and obtain
\begin{eqnarray}
\dot{L_{c3}}&=&(\chi_{S}^{-1}-\chi_{\pi}^{-1})A_S L_{c2}+\rho_{\pi}A_S^{3}\nabla^{2}
L_{c2}  \ \ (c=1,5)
\label{eqn:c22}
\\
\dot{L_{c2}}&=&(\chi_{\pi}^{-1}-\chi_{S}^{-1})A_S L_{c3}-\rho_{\pi}A_S^{3}\nabla^{2}
L_{c3}  \ \ (c=1,5)
\label{eqn:c3}
\\
\dot{L_{43}}&=&\rho_{S}A_S^{3}\nabla^{2}L_{42}
\label{eqn:43}
\\
\dot{L_{42}}&=&-\rho_{S}A_S^{3}\nabla^{2}L_{43} 
\label{eqn:32} \\
\dot{L_{cd}}&=&0  \ \ (c,d\in \{1,4,5\})
\label{eqn:cd2} 
\end{eqnarray}
(\ref{eqn:43}) and (\ref{eqn:32}) lead to a ferromagnetic spin-wave 
excitation with dispersion $\omega=\rho_{S}A_S^{3}k^{2}$, while 
(\ref{eqn:c22}) and (\ref{eqn:c3}) couple to give two 
collective modes, which are fluctuations toward pSC phase.
Anisotropy of $\chi_{ab}$ between pSC and F phase produces a 
mass in these two modes, 
$\omega=
A_S(\chi^{-1}_{\pi}-\chi_{S}^{-1})+\rho_{\pi}A_S^{3}k^{2} $.
Finally, (\ref{eqn:cd2}) is not related with collective 
modes, because $L_{cd} \ (c,d\in\{1,4,5\})$ does not couple with
$L_{23}$.

Let us comment on the case $\chi_{\pi}=\chi_{S}$, 
where the symmetry is higher. At this point unitary pSC states and 
F states are degenerate, and the system undergoes a first-order 
phase transition. All the low-energy collective modes 
have the same dispersion $\omega=\rho_{\pi}A_{S,\pi}^{3}k^{2}$.

In real materials, the spin-orbit
coupling is not negligible and introduces additional anisotropy for the 
direction of spin and $\vec{d}$-vector.
In Sr$_{2}$RuO$_4$, this anisotropy may fix the direction of $\vec{d}$
as $\vec{d}\parallel \hat{e}_{z}$~\cite{xiy1,xiy2}.
We treat this by adding to the 
Hamiltonian extra terms allowed by symmetry.
The combination $k_{x}\pm{\rm i}k_{y}$ gives rise to a
finite value of $\langle \vec{L}\rangle\parallel\hat{e}_{z}$, and,
hence, with the spin-orbit 
coupling $\lambda\vec{S}\cdot\vec{L}$, we expect terms including
$S_{z}$. The gauge invariance allows
the following extra terms up to the second order in the order parameters.
\begin{equation}
H_{so}=-hS_{z}+h'(S_{z})^{2}-2\kappa \{ 
({\rm Re}\pi_{z})^{2}+({\rm Im}\pi_{z})^{2}\}
\label{eqn:hso}
\end{equation}
where $\kappa$ is positive since 
$\langle\vec{\pi}\rangle \propto \vec{d}\parallel\hat{e}_{z}$.
These terms alter the dispersion (\ref{eqn:disp}) to 
\[
\omega^{\pm}(k)=A_{\pi}
\sqrt{\left(\rho_{S}A_{\pi}^{2}k^{2}+\kappa\right)
\left(\chi_{S}^{-1}-\chi_{\pi}^{-1}+\rho_{\pi}A_{\pi}^{2}k^{2}+\kappa\right)}\pm h
\]
i.e.\ the spin-orbit coupling gives a finite mass to these modes \cite{rs}.
The spin susceptibility is then calculated as 
\begin{equation}
{\rm Im}\chi_{xx}={\rm Im}\chi_{yy}\propto
\sum_{\alpha=\pm}\sum_{s=\pm 1}\delta(\omega-s\omega^{\alpha}(k))
\end{equation}
On the other hand, $S_{z}$ is a constant of motion.
Thus, double peaks would be observable in neutron scattering,
due to the spin-orbit coupling.
The additional terms in $H_{so}$ are 
approximate in the sense that we restrict 
the Hilbert space of $\vec d(\vec k)$ by fixing the $\vec k$-dependence.
Strictly speaking we have to take into account the 
tensorial structure due to the $\vec k$-dependence 
in the quadratic term of the $\pi $-fields, 
which is introduced by the spin-orbit coupling. For the above 
mode 
they lead to a minor modification only, but leave qualitative 
aspects unchanged.

The above SO(5) generators and superspins can be cast into a spinor 
form as in \cite{rkdz,henley}, which is relevant for constructing 
a microscopic theory.
Let us introduce a 4-component spinor
\begin{equation}
{}^{t}{\bf \Psi}_{{\bf p}}=\left(
c_{{\bf p}\uparrow},
c_{{\bf p}\downarrow},
g(\hat{{\bf p}})^{*}c_{-{\bf p}\uparrow}^{\dagger},
g(\hat{{\bf p}})^{*}c_{-{\bf p}\downarrow}^{\dagger}
\right),
\end{equation}
where the superscript ${}^{t}$ denotes 
transposition. 
The redundancy in ${\bf \Psi}_{{\bf p}}$ is expressed as 
${\bf \Psi}_{{\bf p}}^{*}=-g(\hat{{\bf p}})R
{\bf \Psi}_{-{\bf p}}$
with $R=\left(
\begin{array}{cc}
0 & 1 \\ -1 & 0
\end{array}
\right)$.
Anticommutation relations are
\[
\{ {\bf \Psi}_{{\bf p}\alpha}^{\dagger},
{\bf \Psi}_{{\bf p}'\beta}\}=\delta_{\alpha\beta}\delta_{{\bf p}{\bf p}'}, \
\{ {\bf \Psi}_{{\bf p}\alpha}^{\dagger},
{\bf \Psi}_{{\bf p}'\beta}^{\dagger} \}=-
g({\bf p})R_{\alpha\beta}\delta_{{\bf p},-{\bf p}'}. 
\]
We introduce a representation of Clifford algebra 
$\{ \Gamma^{a}, \ \Gamma^{b}\}=2\delta_{ab}, (a,b=1,\cdots,5)$;
\[
\Gamma^{1}=\left(
\begin{array}{cc}
0 & -{\rm i}{\sigma}^{y} \\
{\rm i}\sigma^{y} &  0
\end{array}
\right), 
\Gamma^{(2,3,4)}=\left(
\begin{array}{cc}
\vec{\sigma} &  0 \\
0 & {}^{t}\vec{\sigma} 
\end{array}
\right), 
\Gamma^{5}=\left(
\begin{array}{cc}
0 & {\sigma}^{y} \\
{\sigma}^{y} &  0
\end{array}
\right).
\]
in order to write SO(5) generators and superspins as
\begin{equation}
L_{ab}=\frac{1}{8}\sum_{\bf p}
{\bf \Psi}_{{\bf p}}^{\dagger}
\Gamma^{ab}{\bf \Psi}_{{\bf p}},\  \
n_{ab}=\frac{1}{8}\sum_{\bf p}w_{{\bf p}}
{\bf \Psi}_{{\bf p}}^{\dagger}
\Gamma^{ab}{\bf \Psi}_{{\bf p}},
\end{equation}
where $\Gamma^{ab}=-{\rm i}{[}\Gamma^{a},\Gamma^{b}{]}$.
It is noteworthy that the 5-dimensional vector
$\frac{1}{4}\sum_{\bf p}
{\bf \Psi}_{{\bf p}}^{\dagger}
\Gamma^{a}{\bf \Psi}_{{\bf p}}$, an analog of the 
5-dimensional superspin \cite{rkdz} 
in the HTSC, vanishes
identically.
That is why the 
superspin in our SO(5) theory is not 5-dimensional. 
In the SO(5) theory of the HTSC, the situation is opposite; the 
5-dimensional superspin is 
$\frac{1}{4}\sum_{\bf p}w_{{\bf p}}\Psi^{\dagger}_{{\bf p}}\Gamma^{a}
\Psi_{{\bf p}+{\bf Q}}$
in the notation of \cite{rkdz}, while its 10-dimensional 
counterpart 
$\frac{1}{8}\sum_{\bf p}w_{{\bf p}}\Psi^{\dagger}_{{\bf p}}\Gamma^{ab}
\Psi_{{\bf p}+{\bf Q}}$ vanishes.

In conclusion, we have constructed an SO(5) model unifying
p-wave superconductivity and ferromagnetism, which may apply to Sr$_2$RuO$_4$. 
This model describes a phase transition between pSC and F by varying the 
strength of the symmetry-breaking term.
It also predicts spectra of collective modes, among which
are collective modes corresponding the fluctuation between pSC and
F phase. The actual phase diagram close to the transition point 
should be governed by these fluctuations and will be discussed
elsewhere. It would be interesting if the phase transition in
Sr$_2$RuO$_4$ could be observed by changing external parameters such
as pressure. However, the effect of pressure is less clear in the present
model, compared to SO(5) theory of HTSC, where the doping, i.e. 
change of chemical potential, is directly coupled to one of 
SO(5) generators and gives rise to a superspin-flop transition.

The authors acknowledge K.~Imura, Y.~Maeno, J.~Akimitsu, Y.~Hasegawa, 
T.~Imai, and K.~Kanoda for fruitful discussions. 
This work is supported by Grant-in-Aid for 
COE Research No.~08CE2003
from the Ministry of Education, Science, Culture and Sports of Japan.


\begin{thebibliography}{99}

\bibitem{hv1} G.~R.~Stewart, Rev. Mod. Phys. {\bf 56} 755 (1984).
\bibitem{hv2} P.~A.~Lee {\it et al.},
Comments in Solid State Physics {\bf 12} 99 (1986).
\bibitem{org1} D.~Jerome, in {\it Organic Conductors}, 
ed. J.-P.~Farges (Marcel Dekker, New York, 1994) p.405.
\bibitem{htsc} P.~W.~Anderson, Science {\bf 235} 1196 (1987).
\bibitem{zhang}S.~C.~Zhang, Science {\bf 275} 1089 (1997)
and references therein.
\bibitem{demler} E.~Demler and S.~C.~Zhang, 
Phys. Rev. Lett. {\bf 75} 4126 (1995).
\bibitem{exp1} J.~Rossat-Mignot {\it et al.}, Physica C {\bf 185} 86 (1991).
\bibitem{exp2} H.~Mook {\it et al.}, Phys. Rev. Lett. {\bf 70} 3490 (1994).
\bibitem{exp3} H.~F.~Fong {\it et al.}, Phys. Rev. Lett. {\bf 75} 316 (1995).
\bibitem{exp4} H.~F.~Fong {\it et al.}, Phys. Rev. Lett. {\bf 78} 713 (1997).
\bibitem{exp5} P.~Dai {\it et al.}, Phys. Rev. Lett. {\bf 77} 5425 (1996).
\bibitem{mhdz}
S.~Meixner, W.~Hanke, E.~Demler and S.~C.~Zhang, Phys. Rev. Lett.
{\bf 79} 4902 (1997).
\bibitem{ehz}
R.~Eder, W.~Hanke and S.~C.~Zhang, cond-mat/9707233.
\bibitem{headmz}
W.~Hanke {\it et al.}, cond-mat/9807015.
\bibitem{leggett}
A.~J.~Leggett, Rev. Mod. Phys. {\bf 47} 331 (1975).
\bibitem{maeno} Y.~Maeno {\it et al.}, Nature {\bf 372} 532 (1994).
\bibitem{ishida1}
K.~Ishida {\it et al.}, Phys. Rev. {\bf B56} R505 (1997).
\bibitem{mackenzie}
A.~P.~Mackenzie {\it et al.}, Phys. Rev. Lett. {\bf 80} 161 (1998).
\bibitem{ishida2}
K.~Ishida {\it et al.}, to be published in Nature.
\bibitem{rs}
T.~M.~Rice and M.~Sigrist, J. Phys.: Condens. Matter {\bf 7} L643 (1995).
\bibitem{xiy1}
M.~Sigrist {\it et al.}, preprint (1998).
\bibitem{xiy2}
G.~M.~Luke {\it et al.}, Nature {\bf 394} 558 (1998).
\bibitem{ferro} 
T.~Imai {\it et al.}, Phys. Rev. Lett. {\bf 81} 3006 (1998).
\bibitem{yoshida}
K.~Yoshida {\it et al.}, Phys. Rev. {\bf B58} 15062 (1998).
\bibitem{gibb}
T.~C.~Gibb {\it et al.},
J. Solid State Chem. {\bf 11} 17 (1974).
\bibitem{neumeier}
J.~J.~Neumeier, A.~L.~Cornelius and J.~S.~Schilling, Physica B{\bf 198} 
324 (1994).
\bibitem{cao}
G.~Cao {\it et al.}, Phys. Rev. {\bf B55} R5740 (1997).
\bibitem{leggett1}
A.~J.~Leggett, Ann. Phys.{\bf  85} 11 (1974).
\bibitem{ha1}
Y.~Hasegawa, T.~Usagawa and F.~Iwamoto, Prog. Theor. Phys. 
{\bf 62} 1458 (1979).
\bibitem{ha2}
Y.~Hasegawa and H.~Namaizawa, Prog. Theor. Phys. {\bf 67} 389 (1982).
\bibitem{multi}
For a set of more than one pair wavefunction, we should deal with 
the relevant orbital basis functions for $ \hat{{\bf k}} $. 
We consider here the two examples:
(a) $\hat{{\bf k}}=\hat{e}_{x}, \hat{e}_{y}$,  and 
(b) $\hat{{\bf k}}=\hat{e}_{x}, \hat{e}_{y}, \hat{e}_{z}$.
In the case (a), 
the whole algebra decouples into a direct 
product of two SO(5). The generators $L_{ab}^{\pm}$ are 
expressed in a matrix form (\ref{eqn:L}) with
\begin{eqnarray*}
\pi_{\mu}^{\pm}&=&(1/8)\sum_{{\bf p},\alpha}
(\mbox{sgn}({\bf p}_{x})\pm\mbox{sgn}({\bf p}_{y}))
c_{{\bf p}\alpha}
({\sigma}^{y} {\sigma}^{\mu})_{\alpha\beta}c_{-{\bf p}\beta},\\
Q^{\pm}&=&(1/4) \sum_{{\bf p},\alpha} 
(1\pm\mbox{sgn}({\bf p}_{x} {\bf p}_{y}))
\left( c_{{\bf p}\alpha}^{\dagger}
c_{{\bf p}\alpha}-1 /2\right), \\
S_{\mu}^{\pm}&=&(1 / 4)\sum_{{\bf p},\alpha,\beta} 
(1\pm\mbox{sgn}({\bf p}_{x}{\bf p}_{y}))
c_{{\bf p}\alpha}^{\dagger}
{\sigma}^{\mu}_{\alpha\beta}c_{{\bf p}\beta}.
\end{eqnarray*}
The algebra decouples into two SO(5) algebras
corresponding to ordering along the $(1,\pm 1)$-directions.
In (b), the algebra decouples into a direct product of four SO(5) algebras, each 
of which corresponds to ordering along
$(1,1,1)$, 
$(1,-1,-1)$, 
$(-1,1,-1)$ and  $(-1,-1,1)$. 
\bibitem{ab}
P.~W.~Anderson and W.~F.~Brinkman, Phys. Rev. Lett. {\bf 30} 
1108 (1973);  in Proc. of the 15th Scottish Universities Summer School in 
Physics, 1974, edited by J.~G.~M.~Armitage and I.~E.~Farquhar
(Academic Press, New York, 1975) pp.315-416. 
\bibitem{rkdz}
S.~Rabello, H.~Kohno, E.~Demler, S.~C.~Zhang, Phys. Rev. Lett. {\bf 80} 3586 
(1998).
\bibitem{henley}
C.~L.~Henley, Phys. Rev. Lett. {\bf 80} 3590 (1998).
\end{thebibliography}
\end{document}